# Are the Spatial Concentrations of Core-City and Suburban Poverty Converging in the Rust Belt?

Scott W. Hegerty
Department of Economics
Northeastern Illinois University
Chicago, IL 60625
S-Hegerty@neiu.edu

## ABSTRACT

Decades of deindustrialization have led to economic decline and population loss throughout the U.S. Midwest, with the highest national poverty rates found in Detroit, Cleveland, and Buffalo. This poverty is often confined to core cities themselves, however, as many of their surrounding suburbs continue to prosper. Poverty can therefore be highly concentrated at the MSA level, but more evenly distributed within the borders of the city proper. One result of this disparity is that if suburbanites consider poverty to be confined to the central city, they might be less willing to devote resources to alleviate it. But due to recent increases in suburban poverty, particularly since the 2008 recession, such urban-suburban gaps might be shrinking. Using Census tract-level data, this study quantifies poverty concentrations for four "Rust Belt" MSAs, comparing core-city and suburban concentrations in 2000, 2010, and 2015. There is evidence of a large gap between core cities and outlying areas, which is closing in the three highest-poverty cities, but not in Milwaukee. A set of four comparison cities show a smaller, more stable city-suburban divide in the U.S. "Sunbelt," while Chicago resembles a "Rust Belt" metro.

# 1    Introduction

Among large cities in the United States, Detroit, Cleveland, and Buffalo had the three highest poverty rates in 2018.[1] The forces of economic change and globalization have had profound effects on these and other "Rust Belt" U.S. cities, which have seen manufacturing move out in search of lower wages and production costs in other states or abroad. While cities nationwide have also struggled to define new, post-industrial identities, the Northeast and Midwest have experienced particularly drastic urban population losses and decreases in their economic bases. This, in turn, has led to sharp increases in poverty in many of the region's largest cities. But, while many central cities have suffered since the 1950s, many of their suburbs have continued to grow, leading to widening disparities in the distribution of poverty across metropolitan areas. "Detroit," for example, is known for concentrated poverty when the entire metropolitan area is considered, while poverty in the city itself is far more widespread. As a result, metropolitan areas are often highly polarized, with fiscally conservative suburbanites unwilling to pay for programs that reduce poverty in seemingly faraway jurisdictions.

Although poverty is often considered to be primarily an urban problem, suburban poverty has become increasingly widespread in recent decades. Kneebone and Berube (2014) and Allard (2015) extensively study nationwide trends and analyze their underlying causes; in particular, they note regional differences in the geography of urban and suburban poverty. Changes in population, including their numbers, education levels, and racial/ethnic makeup, are key: Murphy and Allard (2015) find that population increases in the suburbs, which often include people of low as well as high income levels, can help lead to a larger share of poor residents in these areas. Economic factors are equally important; in particular, the 2008 financial crisis put stress on the housing market, which helped lead to reduced wealth and more suburban poverty. At the same time, young professionals have moved to Rust Belt central cities in recent years in search of lower housing costs and an affordable quality of life, and movements of lower-income residents out of central cities have left a larger share of poor residents in the suburbs. Together, these trends could lead to a "convergence" of the distributions of urban and suburban poverty.

---

[1] Source: U.S. Census Bureau, cities with populations above 250,000.

    This convergence can have an impact on policy, as many anti-poverty organizations, the availability of federal funding, and various amenities are often concentrated in central cities. As poverty continues to become more "suburban," more money might be spent on the problem, but the infrastructure required to effectively solve problems is often lacking. Public transportation, for example, might be difficult to expand regardless of short-term funding, particularly if there is significant suburban opposition. As it becomes more widespread, poverty might become less stigmatized as local religious organizations or nonprofits take on visible roles in the suburbs, and states might take more responsibility in helping to reduce poverty rates—but this is more of a long-term solution. While such services have been improving in recent years, Murphy and Allard (2015) note that they often vary widely by location.

    While almost all metro areas nationwide face similar challenges, this study focuses on the U.S. Rust Belt, which is worthy of particular attention. There are three main justifications for this focus. First, these core cities have experienced a specific type of decline, often based on their lack of a diversified economy and dependence on manufacturing. Many Rust Belt cities often do not have large and prosperous education or government sectors that can drive post-industrial growth, as is the case in Boston or Minneapolis-St. Paul. Even Pittsburgh, known for its once-dominant steel industry, has successfully reinvented itself around education and medicine.[2] The economic improvement of Rust Belt cities would

---

[2] Bartik and Erickcek (2008), for example, note how "Eds & Meds" have transformed regional

have national consequences, and a coherent regional policy is key in this process.

Second, these MSAs often have stark urban/suburban and Black/White divides, making poverty alleviation politically problematic. Lawmakers in Michigan, for example, were recently criticized for proposing work requirements for public assistance that might be more easily waived outside Detroit. A region that fights poverty in a unified fashion might be more successful than one that pursues a fragmented policy, and poverty convergence might help increase support for such a policy.

Finally, the Rust Belt also enjoy an outsized political importance in state and national politics. The voting power of central cities can impact the partisan outcome of a statewide election. Examples include the perennial "swing state" of Michigan as well as Wisconsin, where the Democratic stronghold of Milwaukee borders suburban, and conservative, Waukesha County. Even reliably Republican Indiana voted for Barack Obama in 2008 based on voters in the Northwest part of the state, and Democrats in Ohio rely on votes in Cuyahoga County. Federal elections, therefore, often swing on the relative strength of urban and suburban voters in economically depressed metropolitan areas.

This study examines the poverty convergence process for four Midwestern metropolitan statistical areas (MSAs) as follows. First, it separates MSA-level poverty distributions into core-city and non-core-city distributions. Using these measures, the study visually examines the distributions of poverty in core cities and suburban areas for the years 2000, 2010, and 2015, focusing on median and other quartile values. Next, it calculates four measures of the spatial distribution of poverty for each MSA, as well as its core-city and outlying tracts for the same time period. Similar distributions and measures are also calculated for a set of four comparison MSAs (two of which can be characterized as Midwestern and two as "Sunbelt"), as a preliminary examination into whether

---

economies and led to changes in intrametropolitan disparities.

trends in the four "Rust Belt" MSAs are specific to the region. Finally, high-poverty suburban tracts are compared with other suburban tracts and with high-poverty core-city tracts, to see whether they are similar to one another in terms of a set of socioeconomic characteristics. Overall, while the investigation indeed finds distinct patterns in the Midwest[3], it also demonstrates the potential to be extended for large cities elsewhere.

## 2  An Overview of the Literature

Scholars across disciplines have long sought to understand and address the causes of growing poverty, as well as economic segregation, across the country. Wilson (2008-2009, and others) has long been instrumental in describing the impacts, as well as the causes, of increases in poverty on local communities. Kasarda (1989) analyzes the underlying industrial changes behind these trends, and obstacles to individual social mobility. Individuals' abilities to move to less-distressed neighborhoods, examined by South and Crowder (1997) and Crowder *et al.* (2012), is essential to reducing poverty concentrations. Yet, as this mobility is limited, poverty concentrations—particularly in large cities—will continue to be high.

This study relies on statistical measures of poverty concentration. Much of the previous research in this area either has focused primarily on the MSA level of analysis, or relies on simple methods such as differences in poverty rates (rather than statistical indices or other methods) to quantify the distribution of poverty. In this first group, Adelman and Jaret (1999) examine the underlying determinants of Black and White poverty rates in 112 MSAs. Aliprantis *et al.* (2013) plot poverty distributions and calculate concentrations for 100 large U.S. MSAs and other metropolitan areas near Cleveland, finding Rust Belt regions such as Milwaukee and Detroit to have the highest concentrations of poverty in both 2000 and the period from 2006 to 2010. As Hegerty (2019) notes, an MSA-level analysis often misses the fact that poverty is often widespread within city borders, with relatively little differentiation between richer and poorer city neighborhoods.

---

[3] While located in the Northeastern state of New York, a case can be made that Buffalo is culturally "Midwestern."

In the second group of studies, Madden (1996) examines the ratio of central-city to MSA poverty rates in 181 MSAs and finds that it rose over the study period; economic growth tended to lower MSA, but not central-city, poverty. Madden (2003) notes that the number of "non-poor" had declined in Northeastern and Midwestern central cities, as well as many first-ring suburbs, between 1970 and 1990; suburban poverty, however, did not increase in the South and West. Cook and Marchant (2006) classify high-poverty Census tracts for 264 MSAs as urban "core" or inner- or outer-ring suburbs, finding little evidence of poverty spreading from cities to suburbs, with the notable exception of older Northeastern cities. Holliday and Dwyer (2009) examine suburban poverty in 328 metropolitan areas nationwide, noting that poor tracts were often common outside first-ring suburbs, and that they had ethnic and economic characteristics that differed markedly from high-poverty urban tracts. Kneebone *et al.* (2011) isolate tracts with "extreme poverty" (above 40 percent) across 100 large and 266 small MSAs, noting growth in suburban poverty between 1990 and 2009.

A number of studies improve upon more descriptive methods by calculating formal metrics for economic segregation and the concentration of poverty. Important contributions include Jargowsky (1996), who develops a "neighborhood sorting index" for income. Ades *et al.* (2010) examine eight large Canadian metro areas, plotting a set of inequality measures over five-year periods. Weinberg (2011) calculates Gini coefficients for poverty in U.S. states, as well as large metro areas and places, between 2005 and 2009. Lichter *et al.* (2012) calculate dissimilarity indices for urban, rural, and suburban places, emphasizing the impact on different ethnic groups. Dwyer (2012) uses a set of concentration and clustering methods—some of which are similar to those that are used in the current study—and finds poverty in a group of MSAs to have become less concentrated during the 1990s due to the poor and "near poor" growing closer together as the rich pulled away.

Most recently, Cooke and Denton (2015) find poverty to be closely linked to population density in the 100 largest U.S. metro areas. Kavanagh *et al.* (2016) evaluate four Scottish cities in 2001 and 2011, calculating relative concentration indices for each. Hegerty (2019) finds a negative relationship between poverty rates and poverty concentration in a set of 74 core cities (excluding suburbs), and a general decrease in poverty concentration from 2010 to 2015. In particular, the city of Detroit is shown to have the nation's most diffuse poverty, due to the fact that nearly every neighborhood is poor, a finding that stands in marked contrast to the high Detroit MSA concentration measures mentioned above.

Overall, there remains a need for region-specific, quantitative studies of the spatial distributions of poverty, within the core cities themselves, as well as in the suburbs. Here, such a study is conducted, examining each MSA and its core-city and suburban tracts over time, using a variety of techniques. These include maps and plots of distributions, in addition to a set of four statistical measures. This study also aims to investigate four key research themes.

First, trends are examined in four Rust Belt cities, which had populations between 250,000 and 750,000 in 2010 and have the highest poverty rates in the nation. This case study makes an important contribution the literature investigating a vitally important U.S. region. Second, statistical techniques are applied to investigate trends in the geographic expansion of suburban poverty over time, expanding the use of quantitative methods to investigate this topic. Third, this study investigates whether these results are specific to the Rust Belt, comparing them to a small set of cities and metros elsewhere in the region and the country. Finally, high-poverty suburban tracts are examined to find out whether they appear to be similar to high-poverty core-city tracts in the Rust Belt, or whether poverty might be associated with different socioeconomic characteristics in the suburbs.

Poverty is found to be less concentrated within all four central cities compared to the suburban parts of their metros, and that MSA-level poverty measures are closer to the suburban tracts' values. Overall, suburban poverty has become less concentrated, particularly from 2010 to 2015. These findings stand in contrast to other U.S. regions, particularly in the U.S. Sunbelt. Poverty in core cities has generally become more concentrated, but this finding varies by city and by concentration measure. Of the four Rust Belt cities, only Milwaukee has continued to diverge from its surrounding area, and in many ways seems to be undergoing a different process than the other three main subject cities in this analysis. This paper proceeds as follows. Section 3 describes the statistical approach. Section 4 provides the results, and Section 5 concludes.

# 3 Methodology

## 3.1 Choice of data and study area

As the U.S. cities with the highest poverty rates in 2018, Detroit, Cleveland, and Buffalo are the driving force behind this study. Also in the top ten is Milwaukee, which has a number of similar characteristics to these three cities. While the definition of "Rust Belt" can vary, these are all located on the Great Lakes in the Midwest and Northeast, have lost a substantial share of industry and population, and are known for stark economic and racial inequalities. For purposes of this study, cities such as Toledo are excluded based on size; Pittsburgh and St. Louis fall outside the chosen geographic area; and as is noted below, Chicago's diversified economy and global-city status sets it apart from its geographic neighbors. Likewise, Columbus (Ohio) is a state capital and university center, so it lacks Cleveland's "Rust Belt" heritage. For that reason, it is included alongside Chicago as a comparison city.

Many of the cities studied here have large suburbs. While they are not specifically analyzed here, they themselves might serve as centers of concentrated poverty. Examples include Warren and Dearborn, which had 2018 poverty rates of 35.6% and 28.3%, respectively; these were close to Detroit's value of 36.4%. East Cleveland's poverty rate of 39.7% was higher than Cleveland's (34.6%). On the other hand, Niagara Falls (12.4%) and Cheektowaga (10.6%) had much lower poverty rates than Buffalo (30.3%), and the differences were even starker between Waukesha (10.7%), Brown Deer (10.1%), Wauwatosa (6.9%), and Milwaukee (26.6%). Any type of municipality-based analysis, particularly one that makes a distinction between "first-ring" and outer suburbs, is left for a separate study. Here, the focus is on distributions rather than the locations of socioeconomic characteristics of any individual municipality.

In the main portion of this study, U.S. Census data from 2000 are used, as well as 2010 and 2015 American Community Survey (ACS) 5-year estimates, for tract-level poverty, population, and land area. While the Census definition of household poverty is often criticized because it is based on a 1960s-era definition of three times a family's food costs, and does not take regional differences in cost of living into account, it performs adequately in this study for two key reasons. First, this study focuses on relative distributions of poverty across each area, which should be correlated with other, more comprehensive measures and should be similar regardless of levels. Second, this measure has relatively little variation over time or within MSAs. While the fact that food costs have been falling since 1963 has led to a downward trend in the poverty line that has lowered the number of households officially in poverty, this trend is small over the period of study. Food made up 15.4 percent of the Bureau of Labor Statistics' CPI-U in December 2001, and just over 14 percent of the basket in December 2015. Housing costs also vary little in the current study, since regions with high and low costs are not directly compared.

The study areas used here are the MSAs of Buffalo (Erie and Niagara Counties), Cleveland (Cuyahoga, Geauga, Lake, Lorain, and Medina), Detroit (Lapeer, Livingston, Macomb,

Oakland, St. Clair, and Wayne), and Milwaukee (Milwaukee, Ozaukee, Washington, and Waukesha).[4] For each, the whole area ("Metro") is measured, before separating the core City tracts and the suburban ("Noncity") tracts. Using ArcGIS, core-city tracts are defined by selecting those tracts that have their centroids within the Census-designated place in question. While these four cities have clearly defined borders, and tracts rarely lie both inside and outside the city, this is not the case for all cities in the United States. When necessary, ArcGIS selections are modified based on tracts that appear to have the majority of their land area within the city limits.

## 3.2 Distributions of poverty with areas and sub-areas

Both visual and quantitative methods are used to assess changes in the spatial distributions of poverty in the chosen MSAs. First, each MSA's tract-level poverty rates are mapped using a consistent scale across time, noting any visually apparent spatial patterns in their increases or decreases over time. Next, the distributions of poverty rates are plotted, ordering each set of tracts from low to high poverty rates, for each MSA's core city, outlying areas, and combined region. It is important to note how these distributions, as well as the overall average poverty rates, change over time for each type of tract (particularly in the middle of each distribution). If the median tract's poverty rate in an MSA's suburbs rises, this might help indicate a change in living conditions beyond what an overall average or a maximum value might predict. If this rate becomes closer in value to the core city's median value, this might indicate spatial convergence.

The corresponding quartiles—the 25, 50, and 75 percent values for poverty rates within each MSA, city, and suburban area in this study—are then calculated for each year, noting differences among MSAs over the time period of study for the poorest households as well as for the middle and upper tiers. These can be used to assess whether poverty has increased only in the poorest tracts, or whether average or better-off tracts have also seen increases or decreases. As a more formal measure, four statistical measures of poverty concentration are calculated for each MSA and its core-city and suburban tracts, and plotted for each year. These alternative measures share important similarities, but also differ in ways that can help uncover key time trends in the distribution of poverty. After examining these trends, specific socioeconomic characteristics in high-poverty suburban tracts are compared with other suburban, as well as urban, tracts. These methods are explained in detail below.

---

[4] These definitions can be found in OMB Bulletin No. 17-01, as well as later bulletins.

## *3.3. Poverty concentration measures*

The first of four measures of spatial inequality is the Gini coefficient, which would equal zero if poverty were equally spread out, and one if it were perfectly concentrated. Tracts are placed in order of increasing poverty rates, into five equally-sized groups based on population. Cumulative proportions of households in poverty for each quintile form a Lorenz curve, which is compared against a 45-degree line using simple geometric methods. The ratio of the distance between the curve and the line, relative to the triangle under the line, gives the coefficient. The best-known of the four measures used here, it is commonly found in country studies of economic development to measure income or wealth equality. It is often calculated using software commands or packages, although here it was done following the formula described here.

Following Ades *et al.* (2010) or Hegerty (2019), three additional concentration indices are calculated:

The *Isolation* index, which uses the numbers of households in poverty ($x_i$) and population ($t_i$), with $X$ and $T$ representing citywide totals of each group:

$$IS = \frac{1}{2}\sum_{i=1}^{n}\left|\frac{x_i}{X} - \frac{t_i - x_i}{T - X}\right| \quad (1a)$$

This score aggregates the "gaps" between tracts' shares of the total poor and non-poor residents. If many tracts have substantially more of one group than the other because poverty is concentrated, the absolute values of these differences will be high, and will sum to a large number.

The *Exposure* index, which uses the product of tracts' shares of the total number of households in poverty, multiplied by the poverty rate.

$$xPx = \sum_{i=1}^{n}\left[\frac{x_i}{X}\right]\left[\frac{x_i}{t_i}\right] \quad (1b)$$

If poverty is concentrated, a few high-poverty tracts will have most of the area's poor residents, and the rest will have a small product of the two ratios. The sum will therefore be low when poverty is concentrated, and high when it is "diffuse." This is opposite from the *IS* score.

The *Absolute Concentration* score, which is the only one of the four measures to incorporate space (land area, $A_i$, for each tract), comparing area taken up by households in poverty versus the total population. Tracts are ordered by population from the smallest to largest, and from largest to smallest, and breakpoints are calculated where the cumulative population equals $X$. $T_1$ is the cumulative population sum from 1 to the breakpoint $n_1$, and $T_2$ is the population sum from breakpoint $n_2$ to $n$:

$$ACO = 1 - \frac{\sum_{i=1}^{n}\frac{x_i A_i}{X} - \sum_{i=1}^{n_1}\frac{t_i A_i}{T_1}}{\sum_{i=n_2}^{n}\frac{t_i A_i}{T_2} - \sum_{i=1}^{n_1}\frac{t_i A_i}{T_1}} \quad (1b)$$

Because the left-hand-side component which incorporates households not in poverty, this score decreases if tracts have large shares of the total number of households in poverty, in a large land area, relative to their shares of the total population. The denominator captures the difference in land area between high- and low-population tracts. The *ACO* score is higher, representing a higher poverty concentration, if disproportionately large shares of total poverty, relative to population, can be found in small land areas.

While each of these measures has unique advantages (the Gini is well-known, the *ACO* is the most "spatial," and the *IS* and *xPx* indices capture the interactions between poor and non-poor households), the latter two are often used. Hegerty (2019), for example, focuses on changes in the *IS* scores for 74 core cities from 2010 to 2015. Including all four measures here will allow one to evaluate similarities and differences in their performance. To do so, each of these scores is plotted in 2000, 2010, and 2015, noting whether these values have increased or decreased during this time. One key point of analysis in this study is the presence of any "convergence" between core-city and suburban poverty distribution as scores get closer together or farther apart. It is expected that the rise of suburban poverty will lead to decreases in these areas' concentration scores; perhaps gentrification will increase core-city concentration scores.

## *3.4. Choice of comparison cities*

Because these Rust Belt findings need context to be meaningful, and to isolate any region-specific trends, much of this analysis is repeated for a set of four comparison cities and their metro areas. While any number of comparison cities are possible, a total of four are chosen to parallel the number of Rust Belt MSAs.

While it is possible to find others that also perform well, the four selected here are chosen based in similarities in MSA size to the four Rust Belt metros, as well as on geographic factors.[5]

Two of the selected cities and MSAs are Midwestern: Chicago and Columbus (Ohio). For the purposes of this study, these are not considered to be "Rust Belt" because of their industrial makeup, but their location can help control for regional characteristics. For the former, the Metro area is defined as Cook County and the 11 counties in Illinois that surround it (Boone, DeKalb, DuPage, Grundy, La Salle, Lake, Kankakee, Kane, Kendall, McHenry, and Will). Columbus, which is nearby to and similar in size to Cleveland, includes the counties of Delaware, Fairfield, Franklin, Hocking, Licking, Madison, Morrow, Perry, Pickaway, and Union in its metro area.

Two similar Sunbelt cities are Orlando (with Lake, Orange, Osceola, and Seminole counties) and Austin (with Bastrop, Caldwell, Hays, Travis, and Williamson counties). It is interesting to note how few cities of the manufacturing-heavy "Rust Belt" type can be found in other areas of the country. Orlando has a large service sector, while Austin is both a state capital and known for its cultural capital. These are selected to show whether these areas exhibit similar distributions and convergence patterns, or whether there are distinct patterns in the Rust Belt cities.

*III.5. The characteristics of areas with high suburban poverty*

The final research question involves to what degree the poorest suburban tracts have socioeconomic characteristics that are similar to those of equally poor city tracts, or whether they are more "suburban" in socioeconomic makeup. This study therefore closely examines the poorest suburban tracts in the four Rust Belt MSAs: Those that exceed their core city's 75$^{th}$–percentile poverty thresholds.[6] While these suburban tracts make up relatively small shares of the total population, these appear to have grown in number and population from 2000 to 2015. Then, differences are compared in group means for a set of six socioeconomic variables, all of which are obtained from Census data (2015 ACS 5-year estimates). The poorest suburban tracts are compared with poor core-city tracts, as well as with the rest of the suburban tracts.

---

[5] Any multi-city metropolitan areas (such as Tampa-St. Petersburg or Minneapolis-St. Paul) are excluded.

[6] As is explained below, for Milwaukee, the median value is used.

Overall, evidence of spatial convergence between core-city and suburban poverty is found for three of the four Rust Belt cities, while comparison cities such as Orlando have higher, and stable, suburban poverty rates across all three time periods. Suburban poverty has different characteristics across the Rust Belt, with Milwaukee standing out among the four cities. The results are provided below.

# 4       Results

## 4.1 Spatial analysis and city characteristics

Table 1 provides statistics describing the shares of core city population and manufacturing employment, well as educational attainment, in both the four Rust Belt cities and the four comparison cities. The former have larger manufacturing shares than do the latter. While the decrease (in terms of percentage points) is smallest in Columbus and Orlando, all eight cities have lost manufacturing. With the exceptions of Milwaukee and Orlando, Rust Belt cities make up smaller shares of their MSAs than do the comparison cities. This in part reflects the fact that Sunbelt cities are often "overbounded," and include areas that would be part of separate suburbs in the Northeast.

Three of the four Rust Belt cities have levels of educational attainment that are lower than both the comparison cities and the national average; only Buffalo is close. In addition, there is little evidence of a disproportionate increase relative to the nation as a whole between 2000 and 2015. While the nationwide share of residents above age 25 with bachelor's and graduate/professional degrees increased by 3.0 and 2.4 percentage points, respectively, only Buffalo's growth in the latter group exceeded the U.S. average. Perhaps talk of gentrification-driven "rebirth" in the region's cities best fits this single case.

Figure 1 shows the spatial distribution of poverty for census tracts in 2000, 2010, and 2015. The four-color scheme reflects each city's 2015 quartile values, with white depicting values below 25 percent and black showing values above 75 percent (the poorest quarter). The maps grow darker over time as suburban poverty rates rise to their highest levels. There is a particular increase in the number of tracts in the third quartile (50-75%) in 2015, which suggests an overall spread in poverty around the middle of many of the distributions. There also is an increase in the poorest quartile in tracts located outside Cleveland and Detroit, in Milwaukee County outside the city of Milwaukee, and in Niagara County north of Buffalo. The concentration measures developed here are then used to verify whether this visual evidence is indeed supported empirically.

## 4.2 Statistical distributions of poverty

Poverty rates, ordered by tract from lowest to highest values for each MSA in 2000, 2010, and 2015 are presented in Figure 2. These are further separated into core city and "Non-City" (Suburban), alongside the full-sample "Metro" distributions. In addition, average values for each year are shown as horizontal lines for each distribution. Overall, core-city tracts have the highest average poverty rates and suburban tracts have the lowest. The full-sample average is by definition in between the two, but these averages are closer to the suburban values.

One important finding is that all average poverty rates have risen from 2000 to 2015, with the exception of the cities of Buffalo and Detroit, where they declined slightly. Perhaps a "comeback" is real here, particularly in Buffalo, which is shown below to have decreasing poverty in its least-poor quarter of tracts. Most interesting from these graphs are the "middle" areas of the distributions, matching the maps above, where median city poverty rates are much higher than elsewhere. Over time, all values in this range have been pushed upward, particularly in suburban areas. The core-city tracts have seen similar dispersion, but to a lesser extent. This supports previous studies that find that suburban poverty has been becoming increasingly less concentrated and more widespread.

Figure 3, which presents poverty quartiles in the Rust Belt for each region and year, confirms these trends. The suburban quartiles appear more compressed because these areas' medians and other values are typically much lower than their corresponding core City values. In Cleveland in 2010, for example, half of core-city tracts had poverty rates above 32.4%, and one-fourth had poverty rates above 40.1%. The corresponding Noncity (suburban) values were 7.2% and 13.4%, respectively. But while these thresholds increased for both areas, the increase was larger for the Noncity tracts, which can be assessed visually as well as numerically. The 75% quartile threshold, for example, rose to 47% for City tracts and 22.8% for Noncity tracts. The poorest half also became somewhat less poor; evidence for convergence is strongest in Buffalo and Detroit, where the median and 25% thresholds dropped in the City, but not suburban, tracts. The lowest quartile value, but not the median, fell in Cleveland. On the contrary, all thresholds (including the lowest) rose for both Milwaukee City and suburban tracts, suggesting a lack of convergence that differs from the other metros. While they rose over the study period, poverty quartiles in suburban Milwaukee are particularly low overall, reflecting the area's well-known ethnic and political segregation.[7]

---

[7] An interesting piece of journalism on the topic is "Dividing lines," by C. Gilbert of the *Milwaukee Journal Sentinel* (May 3, 2014).

## 4.3. Calculated poverty scores

The main objective of this study is to quantify these processes to ascertain whether these visual patterns in the spatial distribution of poverty are indeed validated numerically. The measures of poverty concentration are presented in Figure 4. In most cases, the Metro values track the Noncity values closely, reflecting the suburban share of MSA poverty scores. The Gini coefficients and Isolation indices, which lack an explicitly spatial component, behave very similarly over time (with the exception of Milwaukee). The Exposure score, as expected, runs in an opposite direction from the Isolation index. The *ACO* score, which includes land area in its calculation, exhibits similar behavior to the concentration scores, but the Metro and Noncity values are much closer to one another than to the City scores. "Typical" Gini coefficients are around 0.5 for Metro areas, and *IS* scores are about 0.45. Poverty is more dispersed within the central cities, with Gini coefficients and *IS* scores in the 0.25 to 0.30 range. It is important to note that, other than the Gini coefficient, these measures have no traditional distributions from which statistical significance can be inferred. Here, only general declines and increases are observed.

While some scores disagree with the others (such as Milwaukee's *IS* and Cleveland's *xPx* scores), they generally show strong evidence of convergence in the Buffalo and Detroit MSAs, weaker evidence in the Cleveland MSA, and divergence in the Milwaukee MSA. For the three converging MSAs, the Gini coefficients and *IS* and *ACO* scores are rising in the City tracts, reflecting increased poverty concentration. Suburban tract values are falling, showing increased dispersion. The Metro scores fall along with the Non-core-city scores. Buffalo's convergence process seems to be driven by rising city scores, perhaps suggesting that this city alone is experiencing an urban renaissance that could be connected to its level of human capital. But while suburban poverty in the region is indeed becoming more dispersed, the city of Milwaukee also sees greater dispersion, and therefore is the only MSA where core-city and suburban poverty do not converge.

## 4.4 Characteristics of comparison cities

Are these patterns typical for the U.S., or unique to the industrial Midwest? These are examined by mapping two of the four comparison cities in 2015 and constructing concentration indices for all three time periods for all four cities. Future work could create similar measures for all cities where there are sufficient populations and numbers of Census tracts (there were 78 U.S. cities with populations above 250,000 in 2015), but here the visual nature of this analysis calls for investigating a limited number of cities. In addition, while Northeastern cities such as the fairly rectangular City of Buffalo are straightforward candidates for this type of analysis, irregular borders and urban sprawl might make an expanded list of core cities problematic.

For context, maps of two comparison cities are depicted in Figure 5. The city of Chicago and its 12-county MSA exhibit concentrated poverty, as there are low-poverty tracts both in desirable urban neighborhoods and in the suburbs. Orlando, on the other hand, has relatively high-poverty tracts radiating outward in all directions from the city center. It is assumed, then, that Orlando's concentration scores might reflect more dispersion and will be lower for all except the ("inverted") *xPx* "exposure" index.

Figure 6 indeed shows that for Orlando, and to a lesser extent Columbus and Austin, mean poverty rates are similar between core-city and suburban tracts. The former are generally above 20 percent and the latter a bit lower. Chicago's large city-suburban gap, on the other hand, makes the city resemble its Rust Belt neighbors even though the city has a highly diversified economy. Figure 7 shows increasing median poverty rates in the comparison cities, particularly in the Orlando MSA and in suburban

Chicago. The Midwestern cities have larger urban-suburban disparities; median poverty is more than two times higher in Chicago and Columbus than in their suburbs. The differences are much smaller for Orlando, and suburban Austin's poverty quartiles also appear less compressed than do those in the Midwest. The 75% poverty values rise in all cases, showing increases in extreme poverty, but this is also true for the four Rust Belt cities.

Figure 8 shows the time trends for the comparison cities' poverty distribution scores. Scores are particularly low in Orlando, where poverty is more geographically widespread. Further investigation is necessary to determine whether this tourism-centered area is unique, or whether these reflect unique "Sunbelt" characteristics. There is some evidence of the latter hypothesis in the fact that Austin also has core-city and suburban scores that are much closer together throughout. Poverty has not converged recently, because no divergence had previously occurred. The same might be true for Columbus; only the *xPx* score indicates any type of convergence from 2010 to 2015. Chicago, again, looks more like a "Rust Belt" city in the degree of difference between overall City and Noncity values. Changes from 2010 to 2015, however, appear to be more indicative of stabilization rather than convergence. The Rust Belt does, in fact, exhibit a unique set of characteristics that are worthy of further study.

## 5.5. High-poverty suburban tracts: A comparison

Finally, this study addresses similarities and differences between high-poverty suburban tracts and lower-poverty suburban tracts, as well as versus high-poverty core-city tracts. Table 2 shows the growth, in population and overall percentages, of high-poverty tracts in the non-core-city parts of these MSAs. For the four Rust Belt suburban areas, fewer than 8% of the population lived in tracts with poverty rates above the median in 2015, and fewer than 4% lived in tracts that exceed the 75% thresholds. The corresponding values in 2000 were 2% and 1%, respectively. Populations above the median poverty rate have more than doubled in Cleveland and Milwaukee, and more than tripled in Detroit and Buffalo. There were also large increases in the population of suburban tracts above the 75$^{th}$ City percentile. The comparison cities, on the other hand, did not see such increases, except in the case of Columbus. Highlighting the differences between the two sets of cities, the 2015 percentages of non-core-city tracts with poverty rates above the core-city median are much higher for Austin and Orlando than for any other city; Columbus' is above 10 percent as well. Orlando's values are consistently high, at about 25% throughout. Only Chicago's percentage is lower than those of two of the four Rust Belt cities, confirming this region as a "unique" case.

Figure 9 presents mean values for six socioeconomic variables in high-poverty core-city and suburban Rust-Belt tracts in 2015. In addition, the suburban (Noncity) tracts are split into subgroups, with poverty rates above and below their corresponding city's 75$^{th}$-percentile poverty rates. This will allow us to investigate whether the poorest suburban tracts resemble poor parts of the city more than they do richer areas outside the city. Since Milwaukee is the only MSA with no Noncity tracts at or above the central city's 75$^{th}$ percentile value the median (50 percent) value is used to define "high poverty" tracts instead.

In general, average poverty is higher in the poor City tracts, suggesting that, even as it spreads to the suburbs, this problem is relatively more severe in poor areas of the core city. While some results are similar across all four cities—the percentage of renters is much higher in both poor city and poor suburban tracts than in remaining suburban tracts, for example—others vary by city. Buffalo and Cleveland have high suburban vacancy rates, which are as high as or higher than city levels. Milwaukee's and Detroit's rates are much lower. In the Buffalo MSA, poor suburban tracts resemble poor city tracts for all variables, particularly the share of properties with a second mortgage (indicating a degree of financial stress). This is not the case for the other metros, however; but the exact differences vary by location. Cleveland's "poor Noncity" share is relatively low, Detroit's is relatively high, and Milwaukee's is similar to the other suburban tracts.

Milwaukee again stands out from the others. The city's share of residents above age 25 without a high-school diploma is only high in core-city tracts. While Detroit's suburban tracts also have relatively large shares of White residents, this proportion is even larger for Milwaukee. This city's unique characteristics, including its sharp economic and political divides, are worthy of further investigation. Such a study could be useful not only for Wisconsin, but also for any region that might have similar demographics and exhibit similar behavior in the future. In general, the effectiveness of policies that are designed for and traditionally used in inner cities, but applied to similar suburban areas, should be investigated. Regional policy would benefit as a result, particularly a unified approach that pooled resources across jurisdictions to solve a pervasive problem.

## 5 Conclusion

Decades of economic decline have led to large increases in poverty in the U.S. "Rust Belt," where deindustrialization has often hit harder than elsewhere in the country. Much of this poverty has been concentrated in large inner cities such as Buffalo, Cleveland, Detroit, and Milwaukee, while their surrounding suburbs have often thrived. The result has been concentrated poverty at the metropolitan level. Cities are often "walled off" from their surrounding suburbs (and often from the entire state), leaving leaders and residents to deal with economic decline on their own. Politically, states can become "red" or "blue" based largely on the relative share of the urban vote, leading to additional polarization. But at a national level, Michigan, Ohio, and Wisconsin have been among the most sought-after states in recent elections. The concerns of Rust Belt residents have received outsized attention as of late, but the focus has often been placed on suburban rather than urban voters as conditions deteriorate regionwide.

While MSAs are often considered to have "concentrated" poverty that can be found only in the central city, changing the unit of analysis shows a different picture. Detroit, for example, has been so affected by depopulation and economic decline that urban poverty is ubiquitous within the city limits. Concentration scores can therefore be much lower in the core city compared to the entire metropolitan area. But, given recent increases in suburban poverty, and as gentrification has led in places to an urban revival, it is possible that this urban-suburban divide has narrowed.

This study tests whether convergence has indeed taken place in four Rust Belt MSAs, which some of the country's highest poverty rates, using Census tract data from 2000, 2010, and 2015. Visual depictions, as well as plots of poverty distributions, suggest that this convergence has indeed taken place in most cases. In particular, Buffalo, and Detroit have seen poverty rates in the least-disadvantaged tracts fall, while suburban poverty has spread in all four MSAs. As part of a more rigorous statistical investigation, four measures of poverty concentration are calculated for each MSA, as well as its core-city and suburban tracts, in each year. The concentration of Metropolitan-level poverty has declined for each MSA, and suburban poverty is indeed becoming more widespread. The urban-suburban split has narrowed in all MSAs except Milwaukee, an area with one of the sharpest racial and political divides in the country. A comparison of group means suggests that poor suburban tracts are indeed similar to poor core-city tracts for a set of income, education, and socioeconomic characteristics, but that similarities differ by city and by variable. This is particularly true for Milwaukee, with a large share of white, better-educated suburban poor.

A comparison with four MSAs of similar size to the four Rust Belt MSAs shows much narrower differences between core-city and suburban tracts, particularly in the U.S. Sunbelt. Chicago, on the other hand, exhibits distributions and time trends more like those of its Midwestern neighbors. Most likely, the "Rust Belt" and "Sunbelt" exhibit their own unique patterns of poverty distribution; perhaps all U.S. MSAs could be classified into these, as well as other, categories. Further research—and more space—would be necessary to extend this study to the dozens of large cities in the United States, and in particular, to model the explicit economic processes through which this convergence takes place.

City and regional governments might nonetheless apply two key findings from these results. First, cities nationwide should be careful to adopt policies that are suitable for their specific circumstances, and not rely on policies shown to be effective in a different metropolitan area. Orlando is not Detroit, and vice versa. Second, although cities differ *between* regions such as the Rust Belt and the Sunbelt, similarities *within* regions should be taken into account as poverty concentrations converge—particularly when justifying a unified approach, across jurisdictions, to solve a problem that often seems remote.

Figure 1. Midwestern City and MSA Poverty Rates by Census Tract, 2000-2015.

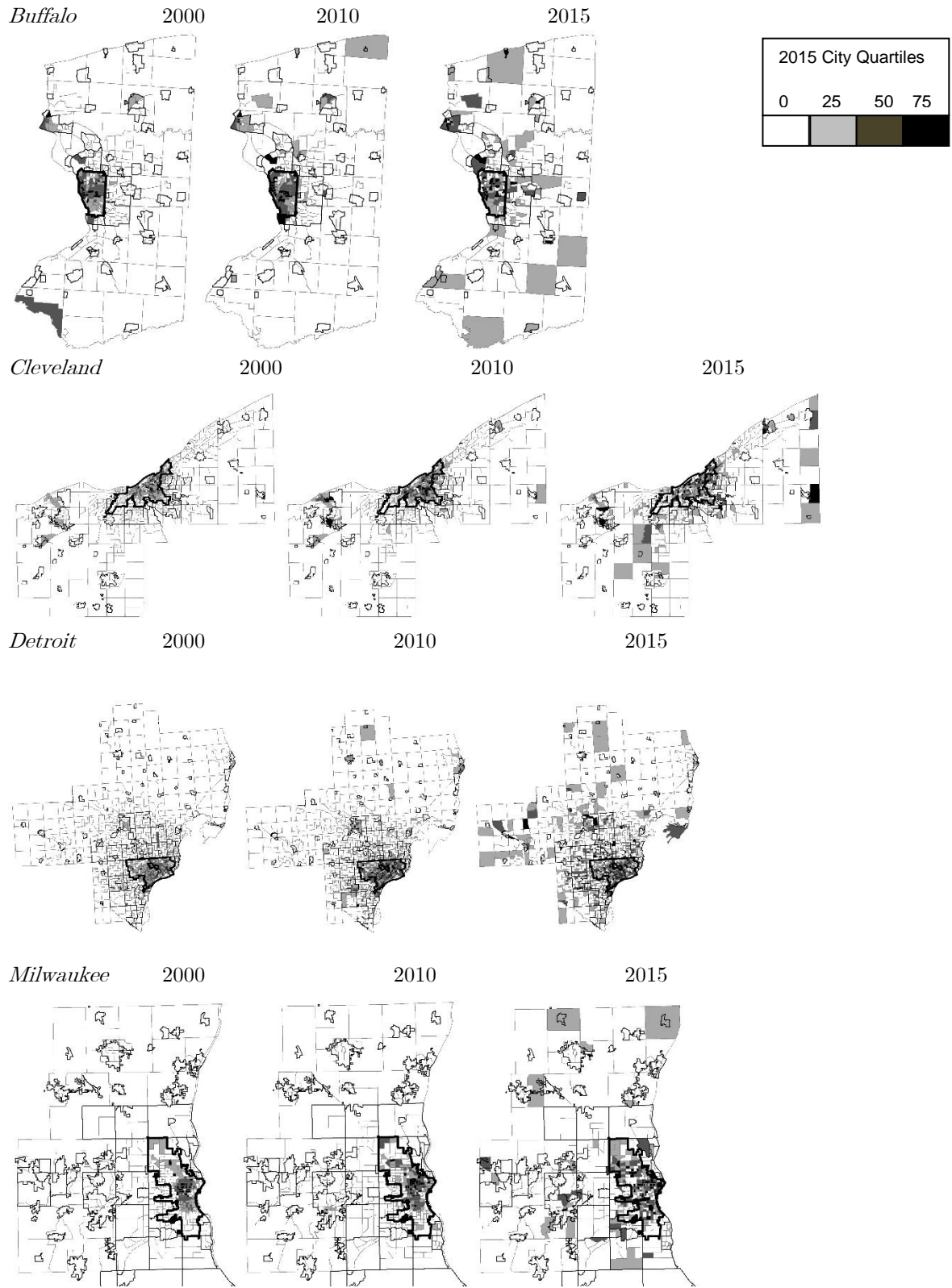

Figure 2. Poverty Rates by Census Tract, Sorted from Lowest to Highest.

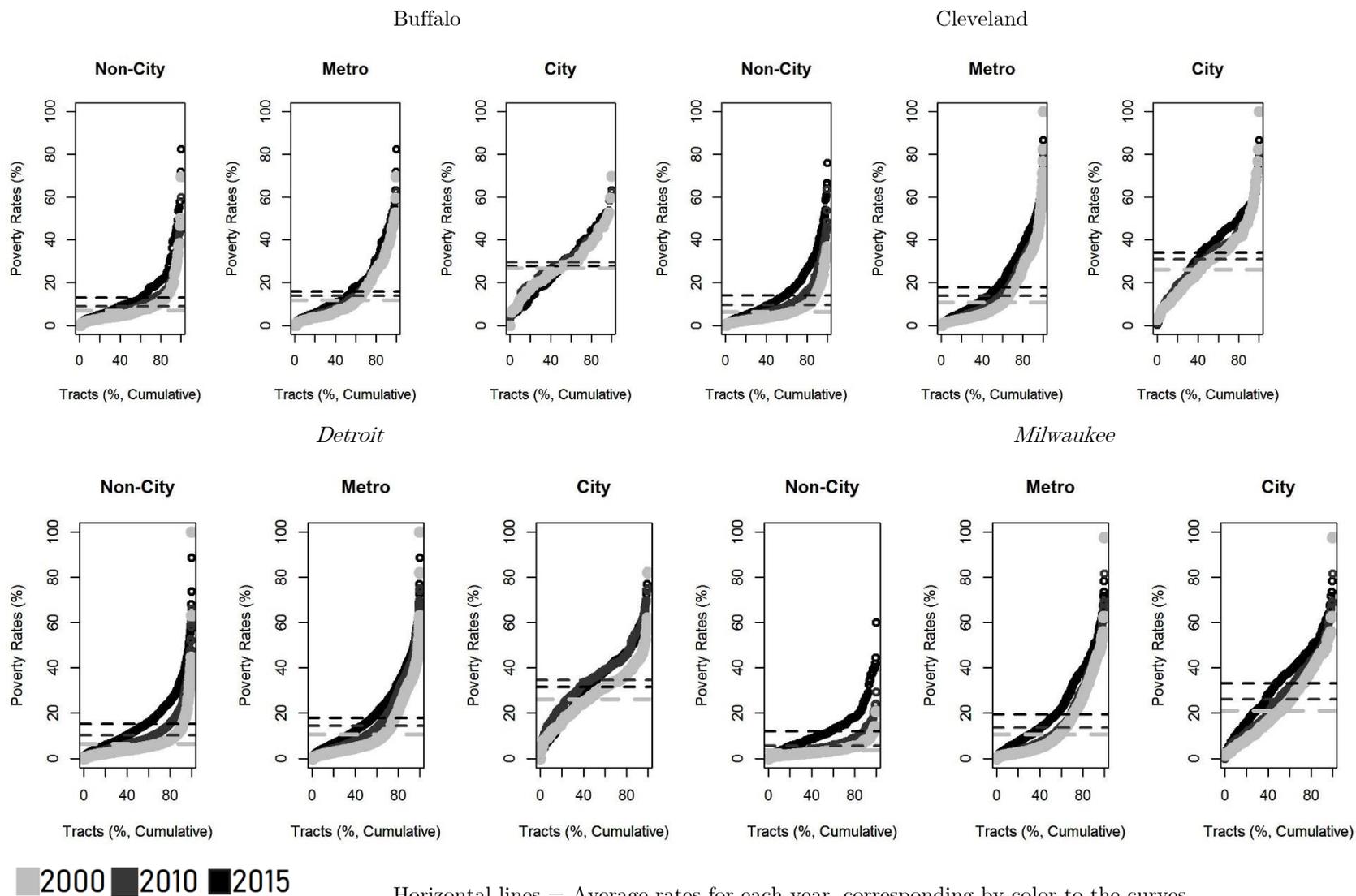

"Non-city" tracts = suburban tracts outside the core city; core-city tracts defined as "City." "MSA" = all tracts for entire defined MSA.

Figure 3. Quartile Values of Core City and Non-City (Suburban) Tract Poverty Rates.

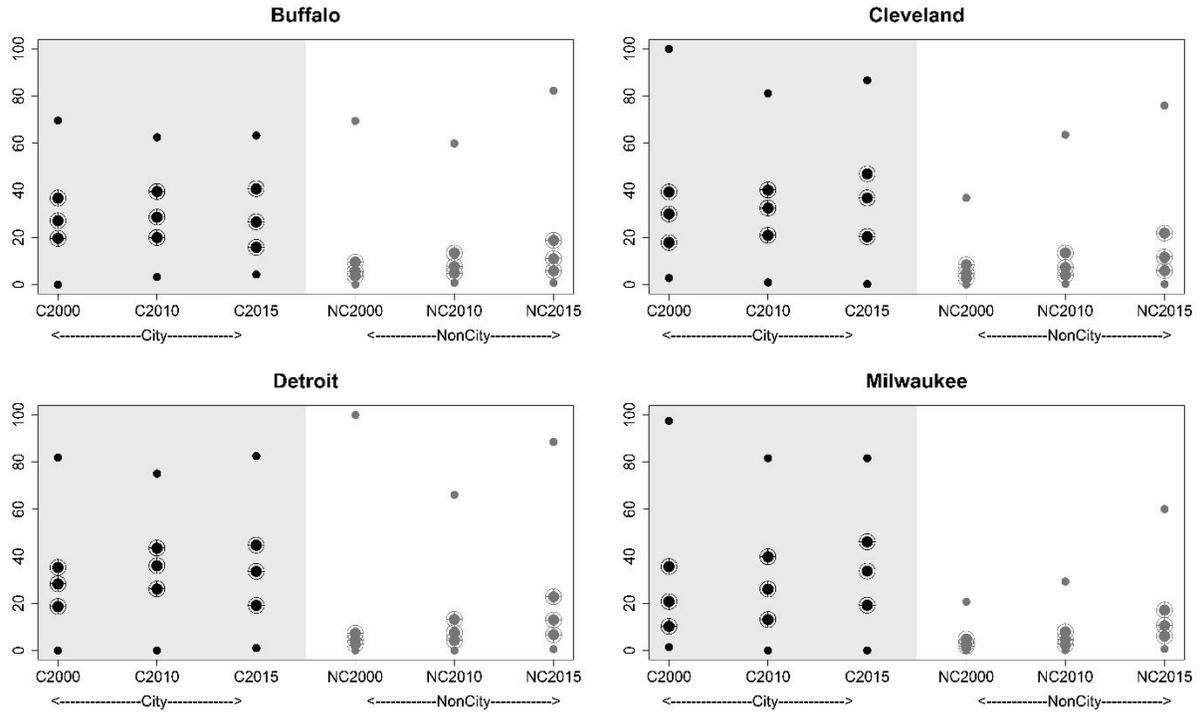

Dots represent 0%, 25%, 50%, 75, and 100% thresholds. 25%, 50%, and 75% thresholds circled.

0

Figure 4. Spatial Measures of Poverty Concentration in 2000, 2010, and 2015.

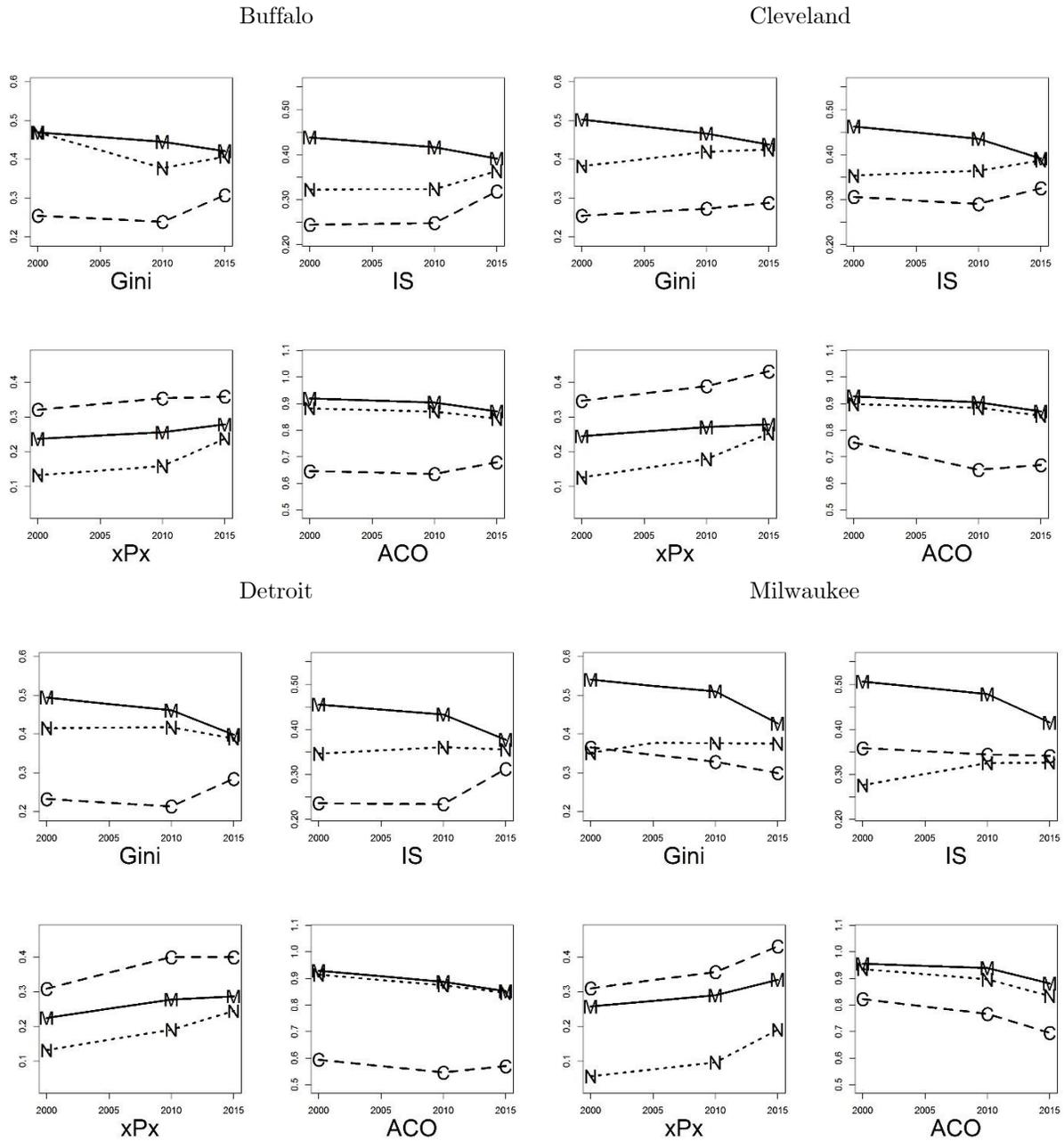

M = Metro (Entire MSA); C = Core City (Within City Limits); N = Non-core-city (Suburban, Remainder of MSA)

Figure 5. Comparison City and MSA Poverty Rates by Census Tract, 2015.

*Chicago (City)*  *(Metro)*  *Orlando (Metro)*

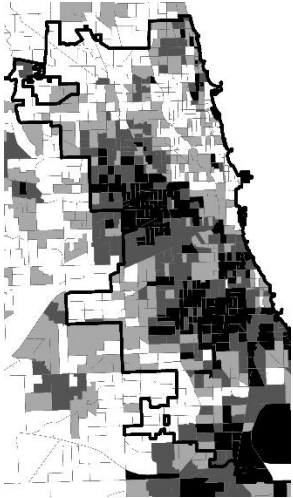
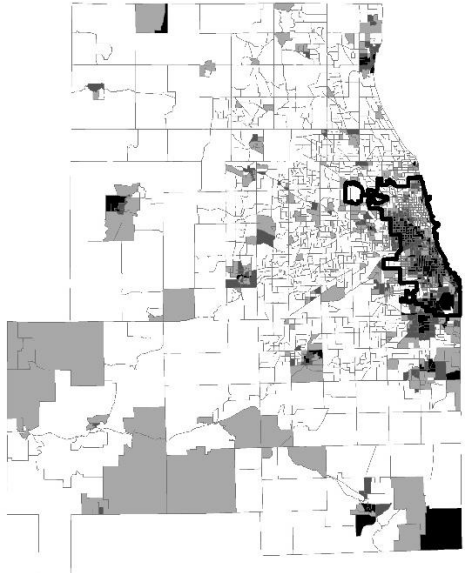
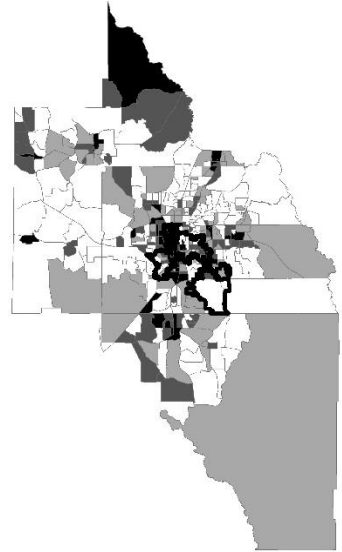

Darker = higher (see Figure 1).

Figure 6. Poverty Rates by Census Tract, Comparison Cities, Sorted from Lowest to Highest.

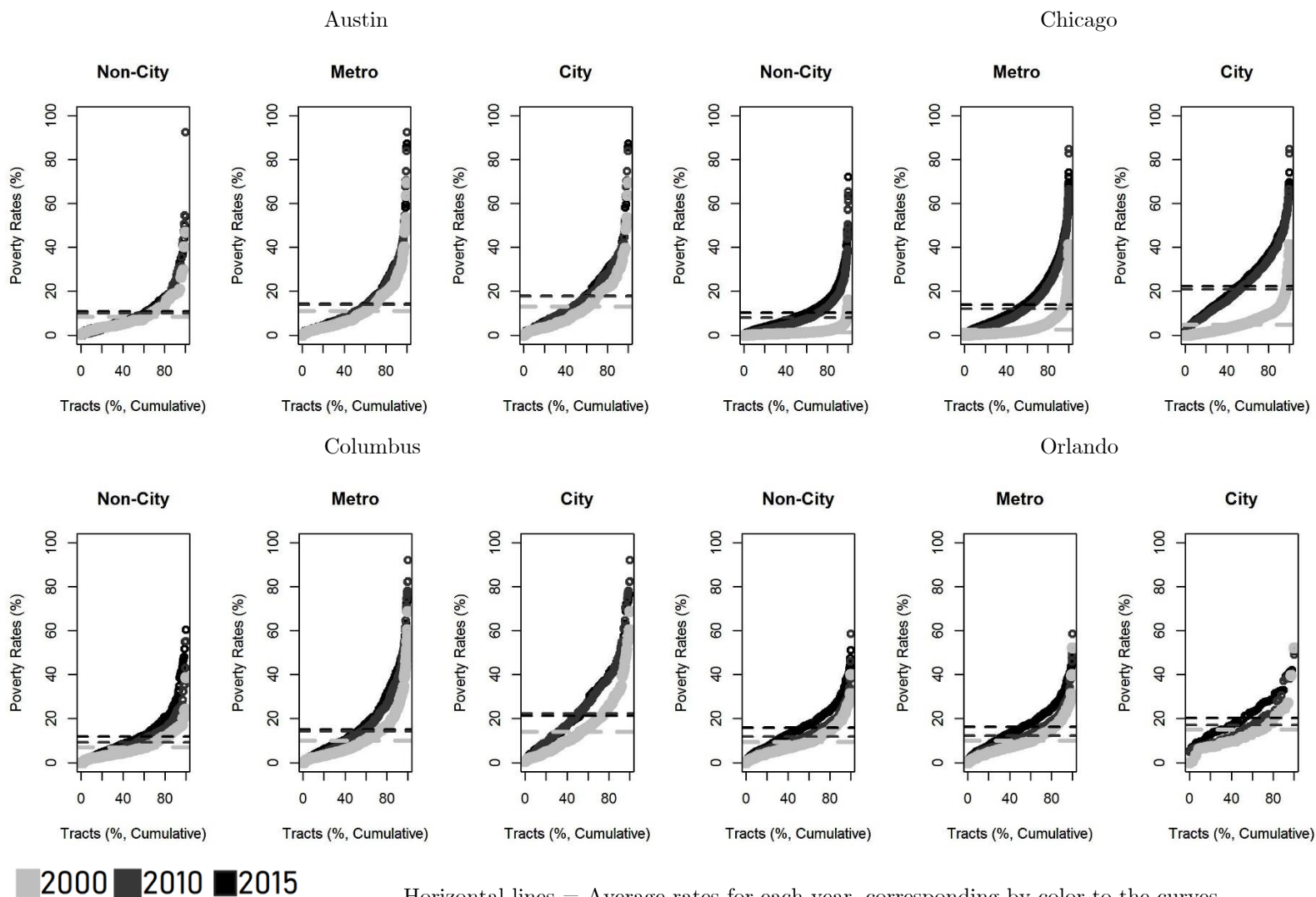

Horizontal lines = Average rates for each year, corresponding by color to the curves.
"Non-city" tracts = suburban tracts outside the core city; core-city tracts defined as "City." "MSA" = all tracts for entire defined MSA.

0

Figure 7. Quartile Values of Comparison City and Non-City Tract Poverty Rates.

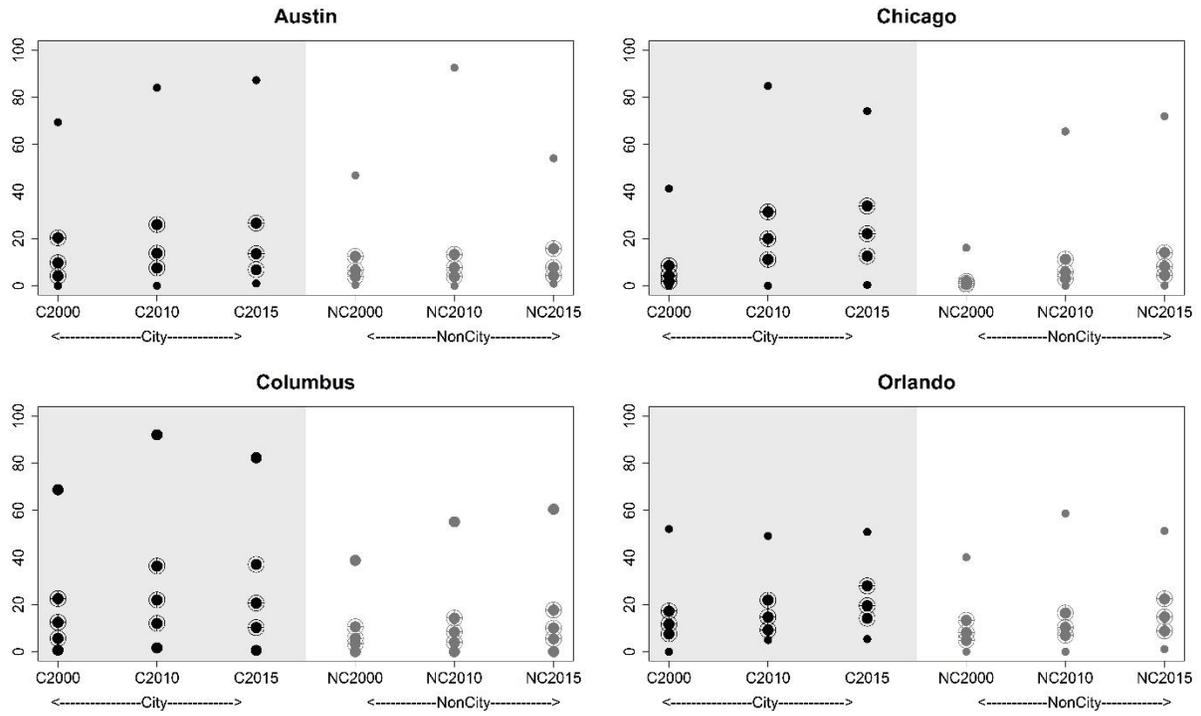

Dots represent 0%, 25%, 50%, 75, and 100% thresholds. 25%, 50%, and 75% thresholds highlighted.

Figure 8. Spatial Measures of Poverty Concentration for Comparison Cities.

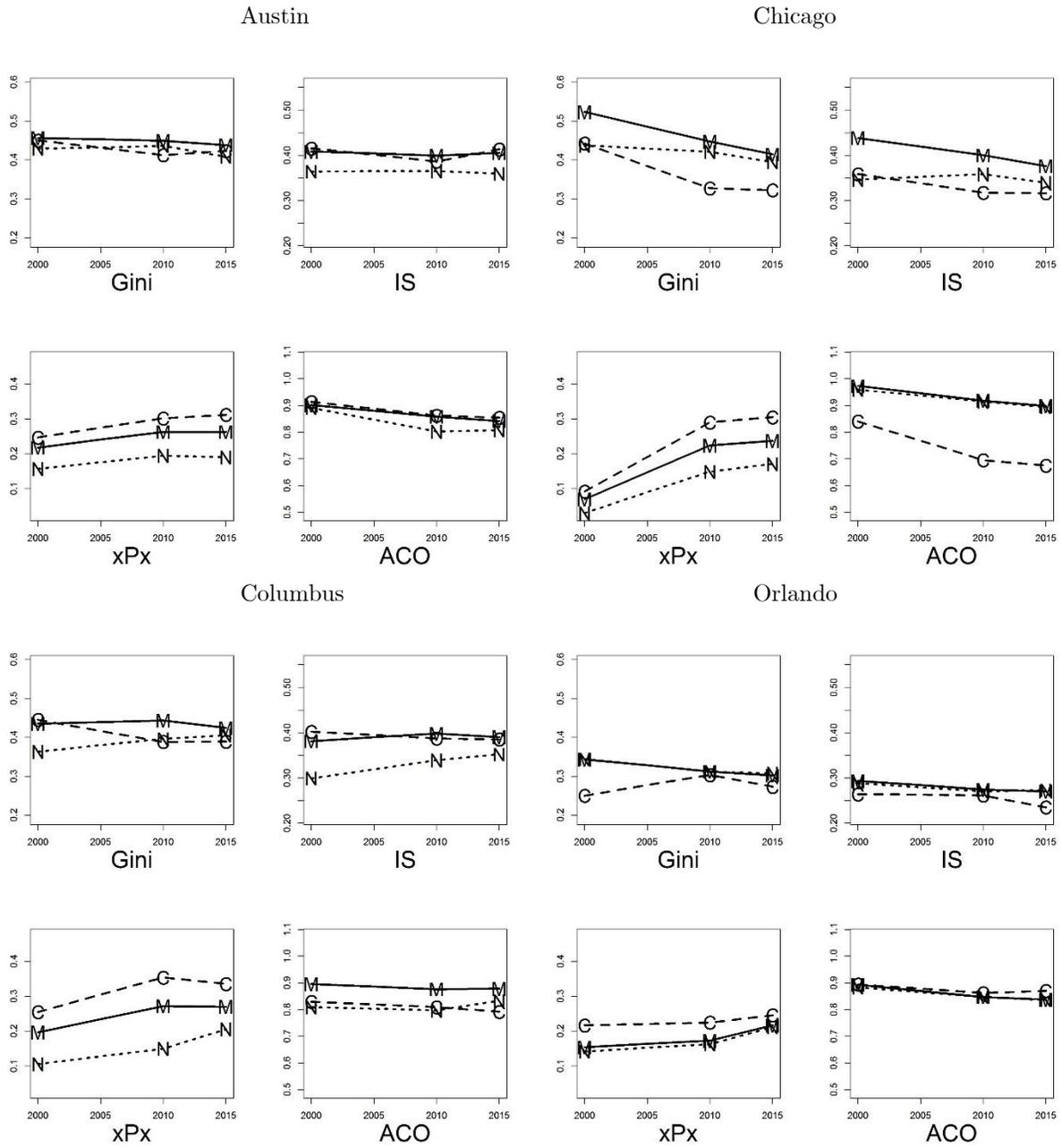

M = Metro (Entire MSA); C = Core City (Within City Limits); N = Non-core-city (Suburban, Remainder of MSA)

Figure 9. Spatial Measures of Poverty Concentration for Comparison Cities.

*Buffalo*

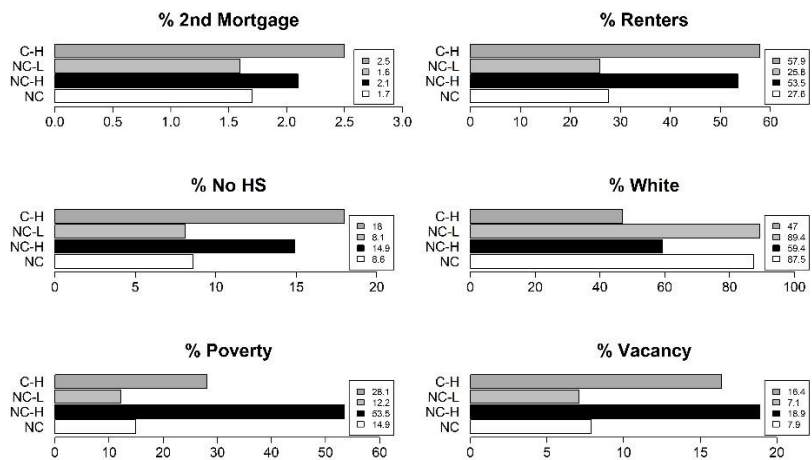

*Cleveland*

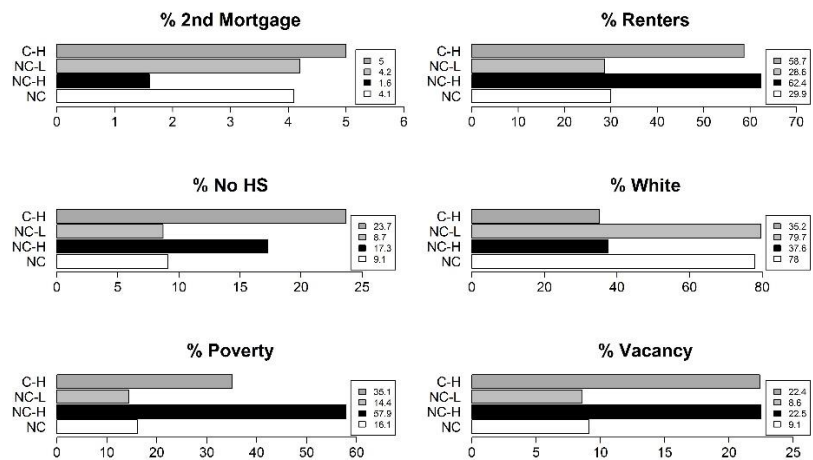

*Detroit*

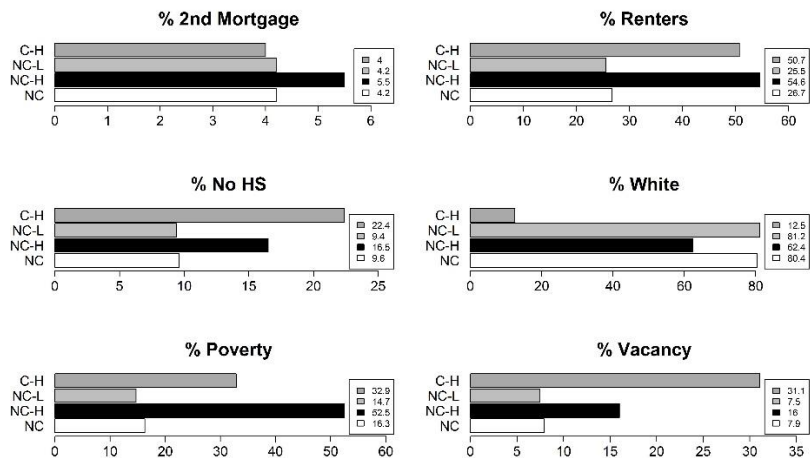

*Milwaukee*

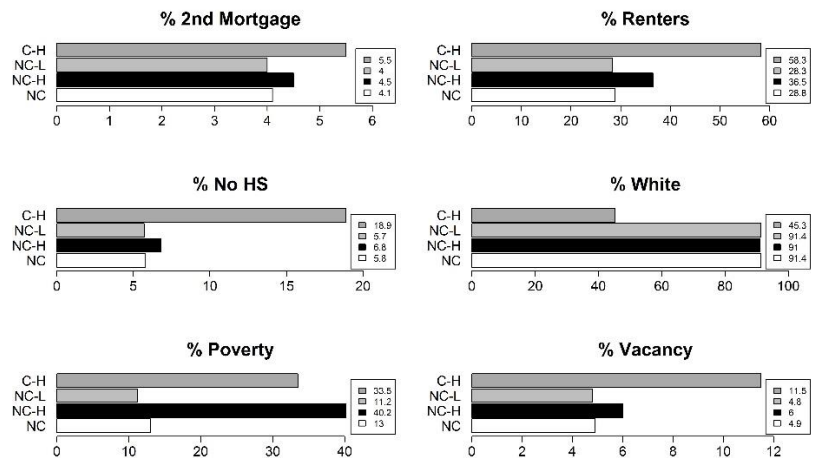

Note: C-H = High-Poverty Core City; NC-L = Low-Poverty Non-Core-City (Suburban); NC-H = High-Poverty Suburban; NC = Entire Non-Core-City (Suburban) area. "High" Threshold = 75% (Threshold for Milwaukee = 50%)

Table 1. Population Statistics for "Rust Belt" and Comparison Core Cities.

| Core City | City Pop. | % Manuf. 2000 | % Manuf. 2015 | % Bach. 2000 | % Bach. 2015 | %Grad. 2000 | % Grad. 2015 | MSA Pop. | % City/MSA |
|---|---|---|---|---|---|---|---|---|---|
| Buffalo | 259517 | 13.1 | 8.9 | 10.5 | 13.4 | 7.8 | 11.2 | 1135734 | 22.9 |
| Cleveland | 390584 | 18.2 | 12.9 | 7.6 | 9.8 | 3.8 | 5.8 | 2064483 | 18.9 |
| Detroit | 690074 | 18.8 | 13.6 | 6.8 | 8.0 | 4.2 | 5.4 | 4296416 | 16.1 |
| Milwaukee | 599498 | 18.5 | 14.3 | 12.3 | 15.4 | 6.0 | 7.9 | 1570006 | 38.2 |
| Austin | 887061 | 12.6 | 7.1 | 25.7 | 29.7 | 14.7 | 17.2 | 1889094 | 47.0 |
| Chicago | 2717534 | 13.1 | 8.9 | 15.5 | 21.2 | 10.0 | 14.3 | 8826671 | 30.8 |
| Columbus | 824663 | 8.9 | 7.4 | 19.9 | 22.3 | 9.2 | 11.9 | 1972375 | 41.8 |
| Orlando | 256738 | 5.6 | 4.0 | 19.9 | 22.8 | 8.3 | 11.5 | 2277816 | 11.3 |

Bach. = Bachelor's degree; national averages = 15.5% (2000) and 22.8% (2015)
Grad. = Graduate or professional degree; national averages = 8.3% (2000) and 11.5% (2015)

Table 2: Non-City Tracts that Exceed the 50th or 75th Percentile for City Poverty.

| Metro Area | Number of tracts 2000 | 2010 | 2015 | Population 2000 | 2010 | 2015 |
|---|---|---|---|---|---|---|
| BFLO (50%) | 11/211 (5.2%) | 11/213 (5.2%) | 27/216 (12.5%) | 23106 (2.03%) | 25902 (2.34%) | 79255 (7.01%) |
| BFLO (75%) | 6/211 (2.8%) | 6/213 (2.8%) | 14/216 (6.5%) | 11269 (0.99%) | 15461 (1.4%) | 40960 (3.62%) |
| CLE (50%) | 13/469 (2.8%) | 25/456 (5.5%) | 44/456 (9.6%) | 31913 (1.51%) | 55765 (2.73%) | 111746 (5.37%) |
| CLE (75%) | 0/469 (0%) | 12/456 (2.6%) | 18/456 (3.9%) | 0 (0%) | 27295 (1.34%) | 41331 (1.99%) |
| DET (50%) | 26/970 (2.7%) | 31/989 (3.1%) | 106/991 (10.7%) | 81680 (1.86%) | 96772 (2.26%) | 334636 (7.57%) |
| DET (75%) | 13/970 (1.3%) | 16/989 (1.6%) | 41/991 (4.1%) | 34301 (0.78%) | 45733 (1.07%) | 106783 (2.42%) |
| MKE (50%) | 0/190 (0%) | 1/217 (0.5%) | 15/219 (6.8%) | 0 (0%) | 2739 (0.18%) | 46400 (3.27%) |
| MKE (75%) | 0/190 (0%) | 0/217 (0%) | 1/219 (0.5%) | 0 (0%) | 0 (0%) | 2814 (0.2%) |
| Metro Area | 2000 | 2010 | 2015 | 2000 | 2010 | 2015 |
| AUS (50%) | 31/92 (33.7%) | 40/164 (24.4%) | 49/166 (33.7%) | 155101 (12.76%) | 173274 (10.88%) | 268941 (14.55%) |
| AUS (75%) | 6/92 (6.5%) | 15/164 (9.1%) | 15/166 (6.5%) | 31325 (2.58%) | 53357 (3.35%) | 69924 (3.78%) |
| CHI (50%) | 65/1069 (6.1%) | 98/1276 (7.7%) | 138/1278 (6.1%) | 189598 (2.93%) | 382468 (4.45%) | 571948 (6.51%) |
| CHI (75%) | 13/1069 (1.2%) | 27/1276 (2.1%) | 39/1278 (1.2%) | 24570 (0.38%) | 87630 (1.02%) | 129328 (1.47%) |
| COL (50%) | 38/218 (17.4%) | 24/242 (9.9%) | 51/241 (17.4%) | 55099 (8.35%) | 80782 (4.45%) | 184373 (10.61%) |
| COL (75%) | 4/216 (1.9%) | 3/242 (1.2%) | 15/241 (1.9%) | 3668 (0.56%) | 4760 (0.26%) | 48895 (2.81%) |
| ORL (50%) | 68/222 (30.6%) | 110/340 (32.4%) | 110/340 (30.6%) | 315409 (24.94%) | 530873 (25.76%) | 598179 (26.7%) |
| ORL (75%) | 30/222 (13.5%) | 37/340 (10.9%) | 44/340 (13.5%) | 136631 (10.8%) | 150443 (7.3%) | 203758 (9.09%) |

Tracts represented as # / Total, with the calculated percentage in parentheses.